\documentclass[12pt]{article}
\usepackage{amsfonts}

\def\sqr#1#2{{\vcenter{\hrule height.#2pt\hbox{\vrule width.#2pt
height#1pt \kern#1pt \vrule width.#2pt}\hrule height.#2pt}}}
\def\square{\mathchoice\sqr64\sqr64\sqr{4.2}3\sqr{3.0}3 \ \!\!\!}

\def\hook{\hbox{\vrule height0pt width4pt depth0.3pt
\vrule height7pt width0.3pt depth0.3pt \vrule height0pt width2pt
depth0pt} }
\def\br{\begin{eqnarray}}
\def\er{\end{eqnarray}}
\def\brn{\begin{eqnarray*}}
\def\ern{\end{eqnarray*}}
\def\er{\end{eqnarray}}
\def\be{\begin{equation}}
\def\ee{\end{equation}}
\def\eqq{&\!\!\!\!\!=\!\!\!\!\!&}
\def\eee{&\!\!\!\!\!\!\!\!\!\!&}
\def\vt{\vartheta}
\def\vp{\varphi}

\def\a{\alpha}
\def\b{\beta}
\def\d{\delta}

\def\la{\lambda}
\def\na{\nabla^2}
\title{{\bf A class of quasi-linear equations in coframe gravity}\\
}
\author{
\thanks {\quad   email itin@sunset.ma.huji.ac.il}
Yakov Itin \\
\footnotesize {Institute of Mathematics}\\
\footnotesize {Hebrew University of Jerusalem}\\
\footnotesize {Givat Ram, Jerusalem 91904, Israel}\\
}
\begin{document}  
\textheight 24cm
\textwidth 16cm        
\oddsidemargin 0.5cm
\evensidemargin 0.5cm
\topmargin=-1.5cm
\footskip 1cm
\parskip=1ex
\newcommand{\sect}[1]{\setcounter{equation}{0}\bigskip\medskip\section{#1}\smallskip}
\newcommand{\subsect}[1]{\medskip\subsection{#1}\smallskip}
\newcommand{\subsubsect}[1]{\medskip\subsubsection{#1}\smallskip}
\renewcommand{\theequation}{\thesection.\arabic{equation}}
\setlength{\evensidemargin}{0in}
\setlength{\oddsidemargin}{0in}
\setlength{\textwidth}{6.25in}
\setlength{\textheight}{8.5in}
\setlength{\topmargin}{0in}
\setlength{\headheight}{0in}
\setlength{\headsep}{0in}
\setlength{\itemsep}{-\parsep}
\setlength{\parskip}{\medskipamount}
\renewcommand{\topfraction}{.9}
\renewcommand{\textfraction}{.1}
\newcommand{\ol}{\setlength{\itemsep}{0pt.}\begin{enumerate}}
\newcommand{\eol}{\end{enumerate}\setlength{\itemsep}{-\parsep}}
\newtheorem{THEOREM}{Theorem}[section]  
\newenvironment{theorem}{\begin{THEOREM} \hspace{-.85em} {\bf :} 
}%
                        {\end{THEOREM}}

\newtheorem{DEFINITION}[THEOREM]{Definition}
\newenvironment{definition}{\begin{DEFINITION} \hspace{-.85em} {\bf 
:} \rm}%
                            {\end{DEFINITION}}

\newtheorem{PROPOSITION}[THEOREM]{Proposition}
\newenvironment{proposition}{\begin{PROPOSITION} \hspace{-.85em} 
{\bf :} }%
                            {\end{PROPOSITION}}
\newcommand{\thm}{\begin{theorem}}

\newcommand{\dfn}{\begin{definition}}

\newcommand{\prf}{\noindent{\bf Proof:} }

\newcommand{\ethm}{\end{theorem}}

\newcommand{\edfn}{\bbox\end{definition}}


\newcommand{\pro}{\begin{proposition}}
\newcommand{\epro}{\end{proposition}}
\newcommand{\eprf}{\bbox}
\newcommand{\beqn}{\begin{equation}}
\newcommand{\eeqn}{\end{equation}}
\newcommand{\wbox}{\mbox{$\sqcap$\llap{$\sqcup$}}}
\newcommand{\bbox}{\vrule height7pt width4pt depth1pt}
\newcommand{\qed}{\bbox}

\newcommand{\bi}[1]{\bibitem{#1}}
\date{}

\maketitle
\begin{abstract}
We have shown recently that the gravity field phenomena can 
be described by a traceless part of the wave-type field equation. 
This is an essentially non-Einsteinian gravity model. It has 
an exact spherically-symmetric static solution, that yields to the 
Yilmaz-Rosen metric. This metric is very close to the Schwarzchild metric.
The wave-type field equation can not be derived from a suitable variational 
principle by free variations, as it was shown by Hehl and his collaborates.
In the present work we are seeking for another field equation having
the same exact spherically-symmetric static solution. 
The differential-geometric 
structure on the manifold endowed with a smooth orthonormal coframe field
is described by the scalar objects of 
anholonomity and its exterior 
derivative. We construct a list of the first 
and second order $SO(1,3)$-covariants (one- and two-indexed quantities) 
and a  quasi-linear 
field equation with free parameters. We fix a part of the parameters   
 by a 
condition that the field equation is satisfied by a quasi-conformal coframe 
with a \it harmonic conformal function.\rm 
Thus we obtain a wide class of field 
equations with a solution that yields  the Majumdar-Papapetrou metric and, 
in particularly, the Yilmaz-Rosen metric, that is viable in the framework of 
three classical tests.  
\end{abstract}
\sect{Introduction}
The teleparallel space was introduced for the first time by 
Einstein \cite{E1}, \cite{EM} in a certain variant of a unified 
theory of gravity and electromagnetism. 
The work of Weitzenb\"{o}ck \cite{We1} was the first 
devoted to the investigation of the geometric structure 
of teleparallel spaces. 
The theories based on this geometrical structure appear in physics for 
time to time in order to give an alternative model of gravity 
or to describe the spin properties of matter. For the recent investigations 
in this area see Refs. \cite{Tr} to
\cite{M-H},  
and \cite {Kawai1} to   
\cite {N-Y}.\\
The investigations in Poincar{\'e} gauge theory and in the metric-affine 
theory of gravity opened a new wide field for applications of the teleparallel 
geometric structure. 
The status of the teleparallel space in the metric-affine 
framework is described in Refs. \cite{hehl95}, \cite{G-H} .\\
In \cite{K-I} a certain type of an alternative 
gravity model based on the teleparallel space was proposed.
We consider a differential $4D$-manifold $M$ endowed with 
a smooth coframe field $\{\vt^a(x), \ a=0,1,2,3\}$. The coframe is 
declared to be pseudo-orthonormal in every point of $M$ 
and this condition defines uniquely 
the following geometrical objects:
\begin{itemize}
\item[{\bf 1.}] a metric on the manifold:
\be
\label{g}
g=\eta_{ab}\vt^a\otimes\vt^b,
\ee
where $\eta_{ab}=diag(1,-1,-1,-1)$ is the Minkowskian metric tensor, 
\item[{\bf 2.}] a Hodge dual map $*\Omega^p\to\Omega^{4-p}$
\be
*\vt^a=\epsilon^{abcd}\vt_b\wedge\vt_c\wedge\vt_d,
\ee
where $\epsilon^{abcd}$ is the completely antisymmetric pseudo-tensor 
and 1-forms  with lowered index is defined to be $\vt_a=\eta_{am}\vt^m$,
\item[{\bf 3.}] in additional to the derivative operator 
$d:\Omega^p\to \Omega^{p+1}$  a dual operator - coderivative is defined as
\be
d^\dagger=-*d*,
\ee
which acts as $d^\dagger:\Omega^p\to \Omega^{p-1}$.
\item [{\bf 4.}]The Hodge - de Rham Laplacian on differential forms (d'Alembertian in 
the case of a manifold with the Lorentzian signature)
\be
\square=d^\dagger d+d d^\dagger 
\ee
\end{itemize}
In \cite{K-I} it was proposed a wave-type field equation:
\be
\label{fe}
\square \ \vt^a=\lambda(x)\vt^a,
\ee
where $\lambda(x)$ is a function of the coframe 
field $\vt^a$ and its derivatives.
Note that for simplicity we use the definitions of the Hodge dual 
and the Hodge - de Rham Laplacian without any sign factor. So we 
restrict ourselves to the case of a fixed dimension and signature. For the 
convenient definitions see Ref. \cite {hehl95}. It can be checked that 
the properties of the coderivative operator and the Laplacian hold also for 
our definition.\\
The equation (\ref{fe}) exhibits a system of hyperbolic second order 
nonlinear PDE. It is a covariant equation  relative to the group of the smooth 
transformations of the local coordinate system on the manifold 
$x^\mu\to \tilde{x}^\mu(x^\mu)$. The additional symmetry is 
an invariance relative to the group of the global (rigid) 
$SO(1,3)$ transformations 
of the coframe field: $\vt^a\to\tilde{\vt}^a={A^a}_b\vt^b$, where 
${A^a}_b \in SO(1,3)$.
Thus, we are interested in a representative class of 
$SO(1,3)$-connected coframes, not in a specific coframe.
 All coframes from the same class 
lead to a unique metric and for the test particles 
to a unique equation of motion. Note that we accept the geodesic principle 
in exactly the same form as it is done in the Einsteinian gravity.
The work of deriving this principle from the field equations of type 
(\ref{fe}) in the 
Einstein-Infeld-Hoffmann manner is in progress.\\
In the case of a static spherically-symmetric space-time the field 
equation (\ref{fe}) has 
a unique spherically-symmetric asymptomatically flat static solution:
\be
\label{Y-R-coframe}
\vt^0=e^{-\frac mr}dt \qquad \vt^i=e^{\frac mr}dx^i \qquad i=1,2,3,
\ee
where $m$ is an arbitrary parameter, which can be interpreted as a mass 
of a central body that produces the field. 
This coframe field leads by (\ref{g}) 
to the Yilmaz-Rosen metric (in isotropic coordinates)
\be
\label{Y-R-metric}
ds^2=e^{-2\frac mr}dt^2-e^{2\frac mr}(dx^2+dy^2+dz^2).
\ee
The metric (\ref{Y-R-metric})  appeared for the first time in 
the Yilmaz scalar model of gravity \cite{yilmaz58} (see also \cite{yilmaz76})
and after a few years in a different framework of Rosen's bimetric 
theory of gravity (see \cite{rose73}, \cite{rose74}). 
The metric (\ref{Y-R-metric}) possesses a following fine 
analytic property: it has the same leading terms in the long-distance 
Taylor expansion as the canonical Schwarzschild metric.
This fact leads to a good accordance 
of the Yilmaz-Rosen metric with the observation data 
(at least in the three classical tests).\\
In \cite{it1} it was shown that a  quasi-conformal coframe conformed 
to (\ref{Y-R-coframe})
\be
\vt^0=e^{-f}dt \qquad \vt^i=e^{f}dx^i \qquad i=1,2,3,
\label{coframe}
\ee
where $f=f(x,y,z)$ is an arbitrary scalar function,
solves the field equation (\ref{fe}) if and only if $f$ is a 
(spatial) harmonic function 
\begin{equation}
\triangle f=f_{xx}+f_{yy}+f_{zz}=0.
\label{f-har}
\end{equation}
By (\ref{g}) this coframe leads  to the metric
\be
ds^2=e^{-2f}dt^2-e^{2f}(dx^2+dy^2+dz^2),
\ee
that is known in the classical theory of GR as the 
Majumdar-Papapetrou metric. Thus, we have for the field equation (\ref{fe})
a wide class of solutions that can be obtained by a specific choice 
of a harmonic function $f$. Some of physical interesting solutions 
of this type are exhibited in \cite{it1}.\\
In the present work we  \it {construct} \rm a  free-parametric class of 
quasi-linear field equations that involves
 in the covariant and $SO(1,3)$-invariant form the coframe 
$\vt^a$ and its first and second derivatives.\\
One of the possibility to fix these numerical free parameters  is
to require the equation to have an exact solution of the form 
(\ref{coframe}) with a harmonic function $f$.\\
 The outline of the work is as following. In the section 2 we develop 
mathematical tools that we need to preserve the covariance and 
$SO(1,3)$-invariance. We consider the 
first order ``derivatives'' of the coframe - 
objects of anholonomity ${C^a}_{bc}$ 
and introduce the second order ``derivatives''
 - $B$-objects  ${B^a}_{bcd}$. 
We reduce the problem of obtaining the diffeomorphic covariant, quasi-linear, 
rigid $SO(1,3)$ invariant equation to purely algebraic problem of 
finding a maximal set of 
two-indexed quantities that involve the $B$-type and  the $C$-type objects.\\
We restrict ourselves to the case of a quasi-linear field equation 
with free parameters 
that is linear in $B$-type objects and quadratic in the $C$-type objects. \\
In the section 3 we study which conditions on the free parameters 
in the field equation should be required in 
order to preserve the solution of the form (\ref{coframe}).
Using the calculations of the second order invariants (Appendix A) 
and considering the leading part of the  field equation, we obtain 
a necessary conditions on the free parameters in order to obtain 
the harmonic equation for the scalar function $f$. These conditions are not 
sufficient because the leading part includes, together with the harmonic terms 
$\triangle f$,  the gradient terms $\nabla^2 f$.
In order to eliminate these gradient  terms we consider a  
quadratic part of the field equation. 
We calculate the quadratic invariants for a pseudo-conformal coframe 
(\ref{coframe}) in Appendix B. Using these calculations we 
obtain a necessary conditions on the parameters that should be 
accepted in order to eliminate the gradient terms in the leading part. 
Thus we obtain a complete system of conditions on the 
free parameters of our quasi-linear field equations. 
This system gives a necessary and sufficient condition for the 
field equation to have a pseudo-conformal solution (\ref{coframe}) with 
a harmonic function $f$.\\
In the section 5 we show that the traceless wave-type field 
equation (\ref{fe}) contains within our general class of the 
quasi-linear field equations. We obtain the expression of the Hodge-de Rham 
Laplacian via the $C$- and $B$-objects. The numerical parameters for the field 
equation (\ref{fe}) are calculated. These parameters satisfy our system 
of harmonic conditions.
\sect{Quasi-linear equations}
In order to \it {construct} \rm some  appropriate field equation one need 
to apply some general principle.   
The principle of the quasi-linearity was formulated for the first time 
by Hehl in the Ref. \cite{Hehl4}. This principle holds in most of the viable 
field theories and it can be used as a heuristic 
tool for construction of new physical models.\\
The action and accordingly the field equations should be linear in the higher 
(second order) derivatives.  We accommodate this principle in the following restricted form: 
the field equation should be \it {linear} \rm in the second order derivatives and 
\it {quadratic} \rm  in the first order derivatives.\\
This restriction can be motivated by the following facts
\begin{itemize}
\item It is necessary to consider the second order derivatives terms 
in the linear form in order to obtain the wave equation as a 
first approximation of the field equation.
\item The quadratic first order derivatives terms are already 
involved in the field equation (\ref{fe}) as well as in the more 
general equation derived from the Rumpf Lagrangian, 
see for instant \cite{Hehl}.
\item Considering the field variable  $\vt^a$ to be dimensionless
one obtain the same dimension  for the linear and quadratic pieces.
As a result all free parameters considering later are dimensionless.
\end{itemize}
In order to construct the covariant and rigid $SO(1,3)$ invariant set of first 
derivatives quantities we begin with the exterior derivative of the 
coframe field $\vt^a$.
This 2-form  can be written uniquely as 
\be
d\vt^a=\frac 12 {C^a}_{bc}\vt^{bc}.
\label{obj}
\ee
We will refer to the coefficients
${C^a}_{bc}$ as \textit{objects of anholonomity}\rm.
These coefficients are antisymmetric in the down indices:
\be
{C^a}_{bc}=-{C^a}_{cb}.
\ee
Using twice the interior product with basis vectors $e_m$ we obtain
\be
{C^a}_{bc}=e_c\hook (e_b \hook d\vt^a).
\ee
The 3-indexed objects ${C^a}_{bc}$ can be contracted in order 
to obtain 1-indexed quantities. Because of the antisymmetry in 
the down indices  only one contraction (up to a sign coefficient) 
of the object of anholonomity is possible:
\be 
C_a={C^m}_{am}=e_m\hook (e_a \hook d\vt^m).
\ee
In order to describe the second derivative terms we consider the 
exterior derivative of the object of of 
anholonomity. 
This  1-form can be expressed by its basic components as  
\be
d{C^a}_{mn}:={B^a}_{mnp}\vt^p.
\label{cur}
\ee
We will refer to the coefficients ${B^a}_{mnp}$ as
 $B$-objects. This is a set of scalar quantities that transforms in 
as a  tensor under the rigid  $SO(1,3)$ pseudo-rotations of the coframe $\vt^a$.  
The explicit expression of the 4-indexed $B$-objects can be written as 
\be
B^a_{mnp}=e_p\hook d{C^a}_{mn}=e_p\hook d\Big (e_n\hook (e_m \hook d\vt^a)\Big).
\ee
Note, that these coefficients are antisymmetric in the middle 
indices:
\be
{B^a}_{mnp}=-{B^a}_{nmp}.
\ee
Contracting two indices in the object ${B^a}_{mnp}$ we obtain 
the following two-indexed $B$-objects 
\be
{}^{(1)}\!B_{ab}:={B_{abm}}^m=e^m\hook d\Big (e_m\hook (e_b \hook d\vt_a)\Big),
\label{defB1}
\ee
\be
{}^{(2)}\!B_{ab}:={B^m}_{mab}=e_b\hook d\Big (e_a\hook (e_m \hook d\vt^m)\Big),
\label{defB2}
\ee
\be
{}^{(3)}\!B_{ab}:={B^m}_{abm}=e_m\hook d\Big (e_b\hook (e_a \hook d\vt^m)\Big).
\label{defB3}
\ee
Note, that
\be
dC_a=d{C^m}_{am}={B^m}_{amp}\vt^p=-{}^{(2)}\!B_{ap}\vt^p.
\ee
Let us apply the contraction of the indices for the 2-indexed $B$-objects.
Note, that $ {}^{(3)}\!B_{ab}$ is antisymmetric and its contraction 
is zero.
We have only one (up to a sign) full contraction 
of the quantities ${B^a}_{nmp}$
\be
B:={{B^a}_{ab}}^b={}^{(1)}\!{B^a}_a={}^{(2)}\!{B^a}_a.
\label{defB}
\ee
We will refer to the quantity $B$ as a scalar $B$-object.\\
Now we are ready to construct a general field equation. 
The coframe field $\vt^a$ has 16 independent components, these components
 in our approach
are the independent dynamical variables. In order to have 
a good defined dynamical system we need to construct a $SO(1,3)$ 
tensorial field equation 
that is a system of 16 independent hyperbolic partial differential 
equations of the second order. Recall that we are looking for 
the field equations, that are
 linear in the $B$-objects and quadratic in the $C$-objects.
The leading (second order) part of the equation 
will be  a linear combination of two-indexed  $B$-objects:
\be
L_{ab}=\b_1{}^{(1)}\!B_{(ab)}+\b_2{}^{(2)}\!B_{(ab)}+\b_3{}^{(3)}\!B_{ab}+
\b_4 \ \!\eta_{ab}B+\b_5{}^{(1)}\!B_{[ab]}+\b_6{}^{(2)}\!B_{[ab]}.
\label{comb1}
\ee
The quadratic part of the equation can be constructed as 
a linear combination of two-indexed terms of the type 
$C\times C$ contracted by the Minkowskian metric $\eta_{ab}$. 
Considering all the possible combination of the indices and 
taking in account the antisymmetry of the objects $C$ we obtain 
the following list of independent two-indexed terms:
\br
\label{A1}
{}^{(1)}\!A_{ab}&:=&C_{abm}C^m \qquad { \ \ \rm general \ matrix}\\
\label{A2}
{}^{(2)}\!A_{ab}&:=&C_{mab}C^m \qquad {\ \ \rm antisymmetric \ matrix}\\
\label{A3}
{}^{(3)}\!A_{ab}&:=&C_{amn}{C_b}^{mn}\qquad{\rm symmetric \ matrix}\\
\label{A4}
{}^{(4)}\!A_{ab}&:=&C_{amn}{{C^m}_b}^n \qquad {\rm general \ matrix}\\
\label{A5}
{}^{(5)}\!A_{ab}&:=&C_{man}{{C^n}_b}^m\qquad{\rm symmetric \ matrix}\\
\label{A6}
{}^{(6)}\!A_{ab}&:=&C_{man}{{C^m}_b}^n\qquad{\rm symmetric \ matrix}\\
\label{A7}
{}^{(7)}\!A_{ab}&:=&C_aC_b\qquad{\ \ \ \  \ \ \ \rm symmetric \ matrix}
\er
Taking the traces of these matrices we obtain the following list of 
scalar type quadratic invariants:
\br
\label{A11}
{}^{(1)}\!A&:=& {}^{(1)}\!\!{A_a}^a=-{}^{(7)}\!\!{A_a}^a\\
\label{A22}
{}^{(2)}\!A&:=& {}^{(3)}\!\!{A_a}^a={}^{(6)}\!\!{A_a}^a\\
\label{A33}
{}^{(3)}\!A&:=& {}^{(4)}\!\!{A_a}^a={}^{(5)}\!\!{A_a}^a
\er
The trace of the antisymmetric matrix ${}^{(2)}\!A_{ab}$ is zero.
Note, that each one of the objects (\ref{A1}-\ref{A33}) is a covariant and 
a $SO(1,3)$ invariant value.\\ 
The quadratic part of the equation can be expressed as a linear 
combination of the values (\ref{A1}-\ref{A33}), namely
\br
\label{comb2}
Q_{ab}&=&\a_1{}^{(1)}\!A_{(ab)}+\a_2{}^{(2)}\!A_{ab}+\a_3{}^{(3)}\!A_{ab}+
\a_4{}^{(4)}\!A_{(ab)}+\a_5{}^{(5)}\!A_{ab}+\nonumber \\
&&\a_6{}^{(6)}\!A_{ab}+\a_7{}^{(7)}\!A_{ab}+\a_8{}^{(1)}\!A_{[ab]}+
\a_9{}^{(4)}\!A_{[ab]}+\nonumber \\
&&
\eta_{ab}\Big(\a_{10}{}^{(1)}\!A+\a_{11}{}^{(2)}\!A+\a_{12}{}^{(3)}\!A\Big)
\er
The general field equation that satisfied the requirement described above  can be expressed as 
\be
\Big(L_{ab}+Q_{ab}\Big)*\vt^b=\kappa \Sigma_a,
\label{equ}
\ee
where $\Sigma_a$ is the energy-momentum current of material fields and $\kappa$ is a couple constant.
We will use the field equation  (\ref{equ}) only in vacuum thus it is enough 
to consider the equation 
\be
L_{ab}+Q_{ab}=0,
\label{equation}
\ee 
where the leading part defined by (\ref{comb1}) and the quadratic part by 
(\ref{comb2}).
\sect{Harmonic conditions}
As we have showed in \cite{K-I} the traceless  wave-type 
field equation  (\ref{fe}) has a unique spherically-symmetric 
asymptomatically flat static solution (\ref{Y-R-coframe}).
This solution includes a scalar function $f=\frac mr$, which is 
a $3D$-harmonic function.
In  \cite{it1} we prove that the equation (\ref{fe}) is satisfied 
by a general quasi-conformal coframe 
$$\vt^0=e^{-f}dt, \qquad \vt^i=e^{f}dx^i, \qquad i=1,2,3,
$$
 where $f=f(x,y,z)$ is an arbitrary $3D$-harmonic function i.e. 
$\triangle f=0$.\\
In this section we are looking for  a general field equation of the 
type  (\ref{equation}) which have the same solution: quasi-conformal coframe 
with an  an arbitrary $3D$-harmonic function.
Let us begin with the leading part of the equation (\ref{equation}). 
The computations in 
the Appendix A for the quasi-conformal coframe (\ref{coframe}) yield
\br
L_{ab}&=&\b_1e^{-2f}
\left(
\begin{array}{cccc}
-\tilde{\triangle} \vp   &0 &0&0\\
0& -\vp_{22}-\vp_{33}   & \vp_{12} &\vp_{13} \\    
0&  \vp_{12} &  -\vp_{11}-\vp_{33}  &\vp_{23} \\
0&  \vp_{13}  &-\vp_{23}  &   -\vp_{11}-\vp_{22} 
\end{array}
\right) \nonumber \\
&&-\b_2e^{-2f}
\left(
\begin{array}{cccc}
0   &0 &0&0\\
0& \vp_{11}   &\vp_{12} &\vp_{13}   \\    
0& \vp_{12}  &  \vp_{22}& \vp_{23}\\
0&  \vp_{13}  &\vp_{23}  & \vp_{33}
\end{array}
\right)+{\b_4}\eta_{ab}e^{-2f}\tilde{\triangle}\vp,
\label{3.1}
\er
where we use  the following notations 
\be
\vp_{ab}:=f_{ab}-f_af_b, \qquad
\tilde{\triangle}\vp:=\sum_{a=1}^3 \vp_{aa}.
\ee
Note, that the expression in the right hand side of (\ref{3.1}) does not involve the parameters $\b_3,\b_5$ and $\b_6$. These parameters remain free. 
In order to eliminate the mixed second order derivative terms 
($f_{ij}$ with  $i \ne j$) 
in the expression (\ref{3.1}), we should take 
\be
\label{cond1}
\b_1=\b_2. 
\ee
Then the leading part of (\ref{equation}) takes the form:
 \be
L_{ab}=- e^{-2f}\tilde{\triangle} \vp 
\left(
\begin{array}{cccc}
(\b_1-\b_4)  &0 &0&0\\
0 &(\b_1+\b_4) &0&0     \\
0 &0  & (\b_1+\b_4)  &0 \\
0 &0  &0 &(\b_1+\b_4)  
\end{array}
\right).
\ee
Note that this form does not yet yield a harmonic equation because 
\be 
\tilde{\triangle} \vp =\triangle f-\nabla^2 f.
\ee
In order to compensate the gradient terms $\nabla^2 f$, we need to add in
the field equation the quadratic part.\\
The computations of the quadratic invariants in 
the Appendix A for the quasi-conformal coframe (\ref{coframe}) yield
\br
Q_{ab}&=&-\a_1e^{-2f}
\left(
\begin{array}{cccc}
\na f   &0 &0&0\\
0&\na-f^2_1 &-f_1f_2&-f_1f_3 \\
0&-f_1f_2  &\na-f^2_2 &-f_2f_3 \\
0& -f_1f_3  &-f_2f_3  &\na-f^2_3
\end{array}
\right)\nonumber \\
&&
+\a_3e^{-2f}\left(
\begin{array}{cccc}
-2\nabla^2 f  &0 &0&0\\
0&  2(\na-f^2_1) &-2f_1f_2& -2f_1f_3\\   
0& -2f_1f_2 & 2(\na-f^2_2) & -2f_2f_3\\
0& -2f_1f_3 & -2f_2f_3 & 2(\na-f^2_3)
\end{array}
\right)\nonumber \\
&&
-\a_4e^{-2f}\left(
\begin{array}{cccc}
\nabla^2 f  &0 &0&0\\
0&  -(\na-f^2_1) &f_1f_2& f_1f_3\\   
0& f_1f_2 & -(\na-f^2_2) & f_2f_3\\
0& f_1f_3 & f_2f_3 & -(\na-f^2_3)
\end{array}
\right)\nonumber \\
&&
+\a_5e^{-2f}\left(
\begin{array}{cccc}
0 &0 &0&0\\
0&  3f_1^2 &3f_1f_2&3 f_1f_3\\   
0& 3f_1f_2 & 3f_2^2 & 3f_2f_3\\
0& 3f_1f_3 & 3f_2f_3 & 3f_3^2
\end{array}
\right)
\nonumber \\
&&+\a_6e^{-2f}\left(
\begin{array}{cccc}
-\nabla^2f &0 &0&0\\
0&2f_1^2+\nabla^2f &2f_1f_2&2 f_1f_3\\   
0& 2f_1f_2 & 2f_2^2+\nabla^2f & 2f_2f_3\\
0& 2f_1f_3 & 2f_2f_3 & 2f_3^2+\nabla^2f
\end{array}
\right)+\nonumber \\
&&
+\a_7e^{-2f}\left(
\begin{array}{cccc}
0 &0 &0&0\\
0&f_1^2 &f_1f_2& f_1f_3\\   
0& f_1f_2 & f_2^2 & f_2f_3\\
0& f_1f_3 & f_2f_3 & f_3^2
\end{array}
\right)
\nonumber \\
&&
+\eta_{ab}\Big(\a_{10}-6\a_{11}-6\a_{12}\Big)e^{-2f}\nabla^2 f.
\er
Recall that the matrix $L_{ab}$ is diagonal 
after our choice (\ref{cond1}) of the coefficients $\a$ .
Thus, our first task is to remove all the off-diagonal terms in the 
matrix $Q_{ab}$. Then we have to take 
\be
\label{cond2}
\a_1-2\a_3-\a_4+3\a_5+2\a_6+\a_7=0.
\ee
In order to compensate the quadratic terms $\na f$ on the diagonal 
of the matrix  $L_{ab}$, we should require
\brn
\b_1-\b_4&=&-\a_1-2\a_3-\a_4-\a_6-\a_{10}-6\a_{11}-6\a_{12},\\
\b_1+\b_4&=&-\a_1+2\a_3+\a_4+\a_6+\a_{10}+6\a_{11}+6\a_{12}.
\ern
Thus we have two additional conditions
\br
\label{cond3}
\b_1&=&-\a_1,\\
\label{cond4}
\b_4&=&2\a_3+\a_4+\a_6+\a_{10}+6\a_{11}+6\a_{12}.
\er
If these two conditions are met, the field equation (\ref{equation}) 
takes on the following form:
\be
L_{ab}+Q_{ab}=- e^{-2f}{\triangle} f
\left(
\begin{array}{cccc}
(\b_1-\b_4)  &0 &0&0\\
0 &(\b_1+\b_4) &0&0     \\
0 &0  & (\b_1+\b_4)  &0 \\
0 &0  &0 &(\b_1+\b_4)  
\end{array}
\right)=0.
\ee
Therefore, in the case when the coefficients $\b_1$ and $\b_4$ do not vanish 
simultaneously, the field equation transforms to the harmonic equation 
$\triangle f=0$.
The result can be stated in the form of the following proposition:
\thm
The field equation 
\be
L_{ab}+Q_{ab}=0,
\ee
where
\be
L_{ab}=\b_1{}^{(1)}\!B_{(ab)}+\b_2{}^{(2)}\!B_{(ab)}+\b_3{}^{(3)}\!B_{ab}+
\b_4\eta_{ab}\!B+\b_5{}^{(1)}\!B_{[ab]}+\b_6{}^{(2)}\!B_{[ab]}
\ee
and
\br
Q_{ab}&=&\a_1{}^{(1)}\!A_{(ab)}+\a_2{}^{(2)}\!A_{ab}+\a_3{}^{(3)}\!A_{ab}+
\a_4{}^{(4)}\!A_{(ab)}+\a_5{}^{(5)}\!A_{ab}+\nonumber \\
&&\a_6{}^{(6)}\!A_{ab}+\a_7{}^{(7)}\!A_{ab}+\a_8{}^{(1)}\!A_{[ab]}+
\a_9{}^{(4)}\!A_{[ab]}+\nonumber \\
&&
\eta_{ab}\Big(\a_{10}{}^{(1)}\!A+\a_{11}{}^{(2)}\!A+\a_{12}{}^{(3)}\!A+
\a_{13}{}^{(4)}\!A\Big),
\er
 is satisfied by a coframe
\be
\vt^0=e^{-f}dx^0, \qquad \vt^i=e^{f}dx^i, \qquad i=1,2,3,
\ee
with a harmonic function $f=f(x,y,z)$ if and only if the following 
conditions hold
\be
\b_1=\b_2, 
\ee
\be
\b_1=-\a_1,
\ee
\be
\a_1-2\a_3-\a_4+3\a_5+2\a_6+\a_7=0,
\ee
\be
\b_4=2\a_3+\a_4+\a_6-\a_{10}+6\a_{11}+6\a_{12}.
\ee
\ethm
\sect{The Hodge-de Rham Laplacian}
Let us  show that the field equation (\ref{fe}) satisfies the conditions of 
the theorem 3.1. 
In order to express  the Hodge-de Rham Laplacian via the $A-$ and 
$B-$objects we need the expressions for the coderivatives of the 
basis 1-forms.
Recall two formulas that are deal with the interior product
\be
\vt^a\wedge(e_a\hook \alpha)=p\alpha,
\label{for1}
\ee
\be
e_a\hook(\vt^a\wedge \alpha)=(4-p)\alpha,
\label{for2}
\ee
where $p=deg (\a)$.
\pro
\be
d^\dagger(\vt^{a_1...a_p})=\frac {1}{p-1}\Big(
\sum_{i=1}^p(-1)^{i}\vt^{a_i}d^\dagger(\vt^{a_1...\hat{a}_i...a_p})+
(-1)^{p}\vt^m\wedge*(d\vt_m\wedge *\vt^{a_1...a_p})
\Big)
\label{co-wedge}
\ee
\epro
\prf
Consider the interior product
\brn
e_m\hook [d^\dagger(\vt^{a_1...a_p})]&=&*(\vt_m\wedge *^2 d*\vt^{a_1...a_p})
=(-1)^{p}*(\vt_m\wedge d*\vt^{a_1...a_p})\\
&=&(-1)^{p}*\Big[-d(\vt_m\wedge *\vt^{a_1...a_p})+
d\vt_m\wedge *\vt^{a_1...a_p}\Big]\\
&=&(-1)^{p}\Big[-d^\dagger[*^2(\vt_m\hook \vt^{a_1...a_p})]+
*(d\vt_m\wedge *\vt^{a_1...a_p})\Big]\\
&=&\sum_{i=1}^p(-1)^{i}\d^{a_i}_md^\dagger\vt^{a_1...\hat{a}_i...a_p}+
(-1)^{p}*(d\vt_m\wedge *\vt^{a_1...a_p}).
\ern
Taking now the exterior product of this expression with 
the form $\vt^m$ and using the formula (\ref{for1})
we obtain the relation (\ref{co-wedge})
\be
(p-1)d^\dagger(\vt^{a_1...a_p})=
\sum_{i=1}^p(-1)^{i}\vt^{a_i}d^\dagger\vt^{a_1...\hat{a}_i...a_p}+
(-1)^{p}\vt^m\wedge*(d\vt_m\wedge *\vt^{a_1...a_p}).
\ee
\eprf\\
Using the formula (\ref{co-wedge}) we express the coderivative of the 
basis forms 
$\vt^a$ via the object anholonomity 
\br
\label{formula1}
d^\dagger(\vt^a)&=&C^a,\\
\label{formula2}
d^\dagger(\vt^{ab})&=&C^a\vt^b-C^b\vt^a-{C_m}^{ab}\vt^m,\\
\label{formula3}
d^\dagger(\vt^{abc})&=&C^a\vt^{bc}-C^b\vt^{ac}+C^c\vt^{ab}+
{C_m}^{bc}\vt^{am}-{C_m}^{ac}\vt^{bm}+{C_m}^{ab}\vt^{cm},\nonumber\\
&&\\
\label{formula4}
d^\dagger(\vt^{abcd})&=&0.
\er
The Hodge-de Rham Laplacian of the basis 1-form 
$$\square \  \vt^a=(dd^\dagger+d^\dagger d)\vt^a$$
can now be expressed as 
\be
\square \ \vt^a=-\Big({}^{(1)}B^a_b+{}^{(2)}{B^a}_b +{C^a}_{bn}C^n-
\frac 12{C^a}_{mn}{C_b}^{mn}\Big)\vt^b.
\label{tria}
\ee
Indeed, using (\ref{formula1})  we obtain
$$
dd^\dagger\vt^a=dC^a=d({C^{ka}}_k)={B^{ka}}_{kb}\vt^b= -{}^{(2)}{B^a}_b\vt^b.
$$
As for the second part of the Laplacian we obtain using (\ref{formula2}) 
\brn
d^\dagger d\vt^a&=&\frac 12 d^\dagger({C^a}_{mn}\vt^{mn})=
\frac 12 *d({C^a}_{mn}*\vt^{mn})\\
&=&\frac 12 \Big({B^a}_{mnk}*(\vt^k\wedge *\vt^{mn}
{C^a}_{mn}d^\dagger\vt^{mn}\Big)\\
&=&\frac 12 \Big({{B^a}_{mn}}^k(\d^m_k\vt^n-\d^n_k\vt^m)
+{C^a}_{mn}(C^m\vt^n-C^n\vt^m -{C_p}^{mn}\vt^p)\Big)\\
&=&{{B^a}_{kn}}^k\vt^n+{C^a}_{mn}C^n\vt^m-\frac 12{C^a}_{mn}{C_p}^{mn}\vt^p\\
&=&(-{}^{(1)}B^a_b +{C^a}_{bn}C^n-\frac 12{C^a}_{mn}{C_b}^{mn})\vt^b.
\ern
Let the 1-form of Laplacian $\square \  \vt^a$ will be written as 
\be
\square \ \vt^a={M^a}_m\vt^m.
\ee 
The field equation (\ref{fe})
$
\square \ \vt^a=\la(x)\vt^a
$
can be rewritten as
$$
{M^a}_m\vt^m=\la(x)\vt^a.
$$
Taking in two sides of this equation the interior product with the basis vector 
$e_m$ we obtain
$${M^a}_b=\la(x)\d^a_b.$$
Taking the trace of matrices we get  $$\la=\frac 14 {M^a}_a$$
and the field equation takes the form
\be
M_{ab}-\frac 14 \eta_{ab}{M^p}_p=0.
\ee
Using the expression (\ref{tria}) we obtain 
\br
&&{}^{(1)}\!{B^a}_b+{}^{(2)}\!{B^a}_b -{C^a}_{bn}C^n-
\frac 12{C^a}_{mn}{C_b}^{mn}-\nonumber\\
&&\frac 14\eta_{ab}
\Big({}^{(1)}\!{B^p}_p+{}^{(2)}\!{B^p}_p -{C^p}_{pn}C^n-
\frac 12{C^p}_{mn}{C_p}^{mn}\Big)=0.
\er
By the definitions of the matrices ${}^{(i)}{A_{ab}}$ we can rewrite 
this equation as
\be
{}^{(1)}\!{B^a}_b+{}^{(2)}\!{B^a}_b-{}^{(1)}\!A_{ab}-
\frac 12 {}^{(3)}\!A_{ab}-
\frac 14\eta_{ab}(2B-{}^{(1)}\!A-\frac 12 {}^{(2)}\!A)=0.
\ee
Thus the non-vanishing coefficients of the general quasi-linear equation 
(\ref{equation}) in this special case take the values
\br
&&\b_1=\b_2=1, \qquad \b_4=-\frac 12, \qquad \a_1=-1,\qquad 
\a_8=1,\nonumber\\
&&\a_3=-\frac 12, \qquad
 \a_{10}=\frac 14, \qquad \a_{11}=\frac 18.
\er
These coefficients satisfy the conditions of the theorem.
Thus we restate from this more general point of view our result \cite{it1} 
that the traceless wave-type field equation (\ref{fe}) has the solution  
(\ref{Y-R-coframe}) with an arbitrary $3D$-harmonic function.
\sect{Concluding remarks }
The field equation we used in \cite{K-I} and \cite{it1} is not 
a viable field equation mostly because it can not be derived from 
a certain action principle without additional constraints. In the same time 
the spherically-symmetric solution of this equation yields to a metric which 
is in a good accordance with three classical tests. \\
We are looking for {\it another field equation} that has the same static 
spherically-symmetric solution. \\
We have constructed a  diffeomorphic covariant quasi-linear field equation 
that involves only the coframe field $\vt^a$ and its first and 
second order derivatives and that is invariant relative to the group of rigid 
$SO(1,3)$ transformations of the coframe field. 
This equation involves a set of free dimensionless numerical parameters.  
We fix a part of these free parameters 
in this field equation  in order to obtain a 
quasi-conformal solution of the equation with a harmonic scalar function.\\
This general solution involves a unique spherically-symmetric solution, 
that leads to the viable (in the framework of three classical tests) 
Yilmaz-Rosen metric. 
Another conditions for fixing the free parameters is a requirement 
for the field equation to be derivable from a suitable action principle.
We study this condition in \cite{itin4}.
\section*{Acknowledgements}
I am deeply grateful to Prof. Hehl and Prof. Kaniel for many helpful 
discussions and valuable comments on the preliminary version of this paper.
\appendix
\sect{Calculation of objects of curvature 
for a pseudo-conformal coframe}
\small{
Let us calculate the objects of anholonomity
for a pseudo-conformal coframe:
\be
\vt^0=e^{-f}dt,\qquad\vt^i=e^fdx^i,\qquad i=1,2,3,\qquad f=f(x,y,z).
\ee
The exterior 
derivatives of the coframe forms are
\be
d\vt^0=e^{-f}(f_1\vt^{01}+f_2\vt^{02}+f_3\vt^{03}),\qquad
d\vt^1=e^{-f}(-f_2\vt^{12}-f_3\vt^{13}),
\ee
\be
d\vt^2=e^{-f}(f_1\vt^{12}-f_3\vt^{23}),\qquad
d\vt^3=e^{-f}(f_1\vt^{13}+f_2\vt^{23}),
\ee
where we use a notation $f_i=\frac{\partial f}{\partial x^i}$.\\
Thus the objects of anholonomity (\ref{obj}) take the following forms
\br
\label{A-C_3}
{C^0}_{mn} \eqq  e^{-f}
\left(
\begin{array}{cccc}
0   &f_1 &f_2 &f_3\\
-f_1& 0   &0    &0   \\
-f_2& 0   &0    &0  \\
-f_3& 0   &0    &0   
\end{array}
\right)\qquad \!\!\!\!\!\!\!
{C^1}_{mn}=e^{-f}
\left(
\begin{array}{cccc}
0   &0    &0    &0    \\
0   &0    &-f_2  &-f_3   \\
0   &f_2 &0    &0     \\
0   &f_3 &0    &0 
\end{array}
\right)\nonumber\\
{C^2}_{mn}\eqq e^{-f}
\left(
\begin{array}{cccc}
0   &0    &0    &0    \\
0   &0    &f_1  &0   \\
0   &-f_1 &0    &-f_3     \\
0   &0    &f_3    &0 
\end{array}
\right)\qquad \!\!\!\!\!\!\!
{C^3}_{mn}=e^{-f}
\left(
\begin{array}{cccc}
0   &0    &0    &0    \\
0   &0    &0    &f_1    \\
0   &0    &0    &f_2     \\
0   &-f_1 &-f_2 &0 
\end{array}
\right)
\er
Let us calculate the quantities $C_a:={C^n}_{an}$
\be
\label{A-C_1}
C_{0}=0, \qquad
C_{1}=e^{-f}f_1,\qquad
C_{2}=e^{-f}f_2,\qquad
C_{3}=e^{-f}f_3.
\ee
The $B$-objects  are defined (\ref{cur}) by the exterior derivative of the 
objects of anholonomity.
Denoting $\vp_{ab}:=f_{ab}-f_af_b$ we have 
\brn
d{C^0}_{mn}\eqq e^{-2f}
\left(
\begin{array}{cccc}
0   &\!\!\vp_{11} &\!\!\vp_{12}&\!\!\vp_{13}\\
\!\!-\vp_{11}&    &    &   \\
\!\!-\vp_{12}&    &\mbox{\Huge 0}   &   \\
\!\!-\vp_{13}&    &    &   
\end{array}
\right)\vt^1
+e^{-2f}
\left(
\begin{array}{cccc}
0   &\!\!\vp_{12} &\!\!\vp_{22}&\!\!\vp_{23}\\
\!\!-\vp_{12}&    &    &   \\
\!\!-\vp_{22}&    &\mbox{\Huge 0}   &   \\
\!\!-\vp_{23}&    &    &   
\end{array}
\right)\vt^2
\ern
$$
+e^{-2f}
\left(
\begin{array}{cccc}
0   &\vp_{13} &\vp_{23}&\vp_{33}\\
-\vp_{13}&    &    &   \\
-\vp_{23}&    &\mbox{\Huge 0}   &   \\
-\vp_{33}&    &    &   
\end{array}
\right)\vt^3\\
$$
Thus the non-vanishing coefficients ${B^0}_{mnp}$ are
\brn
{B^0}_{mn1}\eqq e^{-2f}
\left(
\begin{array}{cccc}
0   &\!\!\vp_{11} &\!\!\vp_{12}&\!\!\vp_{13}\\
\!\!\!\!-\vp_{11}&    &    &   \\
\!\!\!\!-\vp_{12}&    &\mbox{\Huge 0}   &   \\
\!\!\!\!-\vp_{13}&    &    &   
\end{array}
\right),\ \ 
{B^0}_{mn2}\!=\!e^{-2f}
\left(
\begin{array}{cccc}
0   &\!\!\vp_{12} &\!\!\vp_{22}&\!\!\vp_{23}\\
\!\!\!\!-\vp_{12}&    &    &   \\
\!\!\!\!-\vp_{22}&    &\mbox{\Huge 0}   &   \\
\!\!\!\!-\vp_{23}&    &    &   
\end{array}
\right),
\ern
$$
{B^0}_{mn3}=e^{-2f}
\left(
\begin{array}{cccc}
0   &\vp_{13} &\vp_{23}&\vp_{33}\\
-\vp_{13}&    &    &   \\
-\vp_{23}&    &\mbox{\Huge 0}   &   \\
-\vp_{33}&    &    &   
\end{array}
\right)
$$
Accordingly taking the exterior derivatives of 
${C^1}_{mn},{C^1}_{mn},{C^1}_{mn}$ we obtain:
\brn
{B^1}_{mn1}\eqq e^{-2f}
\left(
\begin{array}{cccc}
0   &0 &0&0\\
0&  0  & \!\!\!\!-\vp_{12}   &\!\!\!\! -\vp_{13}  \\
0&  \vp_{12}  & 0  & 0  \\
0&  \vp_{13}  & 0  & 0 
\end{array}
\right),\ \ 
{B^1}_{mn2}=e^{-2f}
\left(
\begin{array}{cccc}
0   &0 &0&0\\
0&  0  & \!\!\!\!-\vp_{22}   & \!\!\!\!-\vp_{23}  \\
0&  \vp_{22}  & 0  & 0  \\
0&  \vp_{23}  & 0  & 0 
\end{array}
\right),
\ern
$$
{B^1}_{mn3}=e^{-2f}
\left(
\begin{array}{cccc}
0   &0 &0&0\\
0&  0  &\!\!\!\! -\vp_{23}   & \!\!\!\!-\vp_{33}  \\
0&  \vp_{23}  & 0  & 0  \\
0&  \vp_{33}  & 0  & 0 
\end{array}
\right).
$$

\brn
{B^2}_{mn1}\eqq e^{-2f}
\left(
\begin{array}{cccc}
0   &0 &0&0\\
0&  0  & \vp_{11}   &0  \\
0&  \!\!\!\!-\vp_{11}  & 0  &\!\!\!\!-\vp_{13} \\
0&  0  &\vp_{13}  & 0  
\end{array}
\right),\ \ 
{B^2}_{mn2}=e^{-2f}
\left(
\begin{array}{cccc}
0   &0 &0&0\\
0&  0  & \vp_{12}   &0  \\
0&  \!\!\!\!-\vp_{12}  & 0  &\!\!\!\!-\vp_{23} \\
0&  0  &\vp_{23}  & 0  
\end{array}
\right),
\ern
$$
{B^2}_{mn3}=e^{-2f}
\left(
\begin{array}{cccc}
0   &0 &0&0\\
0&  0  & \vp_{13}   &0  \\
0&  \!\!\!\!-\vp_{13}  & 0  &\!\!\!\!-\vp_{33} \\
0&  0  &\vp_{33}  & 0  
\end{array}
\right)
$$

\brn
{B^3}_{mn1}\eqq e^{-2f}
\left(
\begin{array}{cccc}
0   &0 &0&0\\
0&  0  & 0&\vp_{11}     \\
0&  0  & 0  &\vp_{12} \\
0&  \!\!\!\!-\vp_{11}  &\!\!\!\!-\vp_{12}  & 0  
\end{array}
\right),\ \
{B^3}_{mn2}=e^{-2f}
\left(
\begin{array}{cccc}
0   &0 &0&0\\
0&  0  & 0&\vp_{12}     \\
0&  0  & 0  &\vp_{22} \\
0&  \!\!\!\!-\vp_{12}  &\!\!\!\!-\vp_{22}  & 0  
\end{array}
\right)
\ern
$$
{B^3}_{mn3}=e^{-2f}
\left(
\begin{array}{cccc}
0   &0 &0&0\\
0&  0  & 0&\vp_{13}     \\
0&  0  & 0  &\vp_{23} \\
0&  \!\!\!\!-\vp_{13}  &\!\!\!\!-\vp_{23}  & 0  
\end{array}
\right)
$$
Let us now calculate the  two-indexed $B$-objects 
{\footnotesize
$B_{ab}$.
\brn
{}^{(1)}\!B_{ab}\eqq{B_{abm}}^m={B_{ab0}}^0+{B_{ab1}}^1+{B_{ab2}}^2+{B_{ab3}}^3\\
\eqq
e^{-2f}
\left(
\begin{array}{cccc}
-\vp_{11}   &0 &0&0\\
0&  0  & \vp_{12} &\vp_{13}     \\  
0&  0  & -\vp_{11} & \\
0&  0  &0  & -\vp_{11}
\end{array}
\right)+e^{-2f}
\left(
\begin{array}{cccc}
-\vp_{22}   &0 &0&0\\
0& -\vp_{22}   & 0&\vp_{13}     \\  
0&  \vp_{12}  & 0  &\vp_{23} \\
0&  0  &-\vp_{23}  &  -\vp_{22}  
\end{array}
\right)\\
\eee+e^{-2f}
\left(
\begin{array}{cccc}
-\vp_{33}   &0 &0&0\\
0& -\vp_{33}   & 0&0   \\    
0&  0  &   -\vp_{33}&0 \\
0&  \vp_{13}  &\vp_{23}  & 0  
\end{array}
\right)
\ern}
Thus we obtain
\be
{}^{(1)}\!B_{ab}=e^{-2f}
\left(
\begin{array}{cccc}
-\tilde{\triangle} \vp   &0 &0&0\\
0& -\vp_{22}-\vp_{33}   & \vp_{12}&\vp_{13}     \\
0&  \vp_{12} &  -\vp_{11}-\vp_{33}  &\vp_{23} \\
0&  \vp_{13}  &-\vp_{23}  &   -\vp_{11}-\vp_{22} 
\end{array}
\right)
\ee
For the object ${}^{(2)}\!B_{ab}$ we obtain
\brn
{}^{(2)}\!B_{ab}\eqq{B^m}_{mab}={B^0}_{0ab}+{B^1}_{1ab}+{B^2}_{2ab}+{B^3}_{3ab}\\
&=&e^{-2f}
\left(
\begin{array}{cccc}
0   &0 &0&0\\
0& \vp_{11}   &\vp_{12} &\vp_{13}   \\    
0& \vp_{12}  &  \vp_{22}& \vp_{23}\\
0&  \vp_{13}  &\vp_{23}  & \vp_{33}
\end{array}
\right)+e^{-2f}
\left(
\begin{array}{cccc}
0   &0 &0&0\\
0& 0   &0 &0   \\                        
0& -\vp_{12}  &  -\vp_{22}& -\vp_{23}\\
0&  -\vp_{13}  &-\vp_{23}  & -\vp_{33}
\end{array}
\right)+\\
\eee \!\!\!\!\!\!\!\!\!\! e^{-2f}
\left(
\begin{array}{cccc}
0   &0 &0&0\\
0& -\vp_{11}   &-\vp_{12} &-\vp_{13}   \\    
0& 0  &  0& 0\\
0&  -\vp_{13}  &-\vp_{23}  & -\vp_{33}
\end{array}
\right)+e^{-2f}
\left(
\begin{array}{cccc}
0   &0 &0&0\\
0& -\vp_{11}   &-\vp_{12} &-\vp_{13}   \\    
0& -\vp_{12}  &  -\vp_{22}& -\vp_{23}\\
0&  0  &0  &0 
\end{array}
\right)
\ern
Therefore 
\be
{}^{(2)}\!B_{ab}=-e^{-2f}
\left(
\begin{array}{cccc}
0   &0 &0&0\\
0& \vp_{11}   &\vp_{12} &\vp_{13}   \\    
0& \vp_{12}  &  \vp_{22}& \vp_{23}\\
0&  \vp_{13}  &\vp_{23}  & \vp_{33}
\end{array}
\right)
\ee
The antisymmetric matrix ${}^{(2)}\!B_{ab}$ is vanished:
\be
{}^{(3)}\!B_{ab}={B^m}_{abm}={B^0}_{ab0}+{B^1}_{ab1}+{B^2}_{ab2}+{B^3}_{ab3}=0
\ee
For the scalar $B$-object we obtain
\be
B={}^{(1)}\!{B^m}_m={}^{(2)}\!{B^m}_m=
  e^{-2f}\tilde{\triangle}\vp
\ee
\sect{Calculation of the quadratic invariants for 
a pseudo-conformal coframe}
\subsection{Calculation of ${}^{(1)}\!A_{ab}$}
Using the values (\ref{A-C_1}) we can rewrite (\ref{A1}) 
as follows
$${}^{(1)}\!A_{ab}:=C_{abm}C^m
=-e^{-f}(C_{ab1}f_1+C_{ab3}f_2+C_{ab3}f_3)
$$
Using the matrices (\ref{A-C_3}) we have
\be
\label{B-A1}
{}^{(1)}\!A_{ab}=-e^{-2f}
\left(
\begin{array}{cccc}
\na f   &0 &0&0\\
0&f^2_2+f^2_3 &-f_1f_2&-f_1f_3 \\
0&-f_1f_2  &f_1^2 +f^2_3 &-f_2f_3 \\
0& -f_1f_3  &-f_2f_3  &f_1^2 +f^2_2
\end{array}
\right),
\ee
where $\nabla^2 f=f_1^2+f_2^2+f_3^2$.\\
Note, that for a  pseudo-conformal coframe the matrix ${}^{(1)}\!A_{ab}$ 
is symmetric, its trace takes the value 
\be
\label{B-A1t}
{}^{(1)}\!A={}^{(1)}{A^a}_a=e^{-2f}\nabla^2 f
\ee
\subsection{Calculation of ${}^{(2)}\!A_{ab}$}
The antisymmetric matrix ${}^{(2)}\!A_{ab}$ vanishes for the pseudo-conformal 
coframe (\ref{coframe}) identical. Indeed 
$$
{}^{(2)}\!A_{ab}={C^m}_{ab}C_m=e^{-f}(f_1{C^1}_{ab}+f_2{C^2}_{ab}+
f_3{C^3}_{ab})
$$
\brn
\label{B-A2}
{}^{(2)}\!A_{ab}\eqq e^{-2f}\left(
\begin{array}{cccc}
0   &0 &0&0\\
0& 0   &-f_1f_2 &-f_1f_3  \\   
0& f_1f_2  &  0& 0\\
0& f_1f_3  & 0 & 0
\end{array}
\right)+e^{-2f}\left(
\begin{array}{cccc}
0   &0 &0&0\\
0& 0  &f_1f_2&0  \\   
0& -f_1f_2  &  0& -f_2f_3\\
0& 0 & f_2f_3 & 0
\end{array}
\right)\\
\eee +e^{-2f}\left(
\begin{array}{cccc}
0   &0 &0&0\\
0& 0  &0& f_1f_3\\   
0& 0 &  0& f_2f_3\\
0& -f_1f_3 & -f_2f_3 & 0
\end{array}
\right)=0
\ern
\subsection{Calculation of ${}^{(3)}\!A_{ab}$}
The symmetric matrix ${}^{(3)}\!A_{ab}$ can be direct calculated as following 
$$
{}^{(3)}\!A_{ab}=C_{amn}{C_b}^{mn}=C_{a00}{C_b}^{00}+C_{a01}{C_b}^{01}+
...+C_{a33}{C_b}^{33}
$$
We obtain
\be
\label{B-A3}
{}^{(3)}\!A_{ab}=e^{-2f}\left(
\begin{array}{cccc}
-2\nabla^2 f  &0 &0&0\\
0&  2(f_2^2+f_3^2) &-2f_1f_2& -2f_1f_3\\   
0& -2f_1f_2 & 2(f_1^2+f_3^2) & -2f_2f_3\\
0& -2f_1f_3 & -2f_2f_3 & 2(f_1^2+f_2^2)
\end{array}
\right)
\ee
The trace of this matrix takes the form
\be
\label{B-A2t}
{}^{(2)}\!A={}^{(3)}{A^a}_a=-6e^{-2f}\nabla^2 f
\ee
\subsection{Calculation of ${}^{(4)}\!A_{ab}$}
The general matrix ${}^{(4)}\!A_{ab}$ can be rewritten as follows
\brn
{}^{(4)}\!A_{ab}&=&C_{amn}{{C^m}_b}^n=-C_{a01}({C^0}_{b1}+{C^1}_{b0})
-C_{a02}({C^0}_{b2}+{C^2}_{b0})\\
&&-C_{a03}({C^0}_{b3}+{C^3}_{b0})
-C_{a12}({C^1}_{b2}+{C^2}_{b1})-C_{a13}({C^1}_{b3}+{C^3}_{b1})\\
&&-C_{a23}({C^2}_{b3}+{C^3}_{b2})
\ern
The direct calculations yield 
\be
\label{B-A4}
{}^{(4)}\!A_{ab}=-e^{-2f}\left(
\begin{array}{cccc}
\nabla^2 f  &0 &0&0\\
0&  -(f_2^2+f_3^2) &f_1f_2& f_1f_3\\   
0& f_1f_2 & -(f_1^2+f_3^2) & f_2f_3\\
0& f_1f_3 & f_2f_3 & -(f_1^2+f_2^2)
\end{array}
\right)
\ee
Note, that for the coframe (\ref{coframe}) the matrix is symmetric 
and its trace takes the value
\be
\label{B-A3t}
{}^{(3)}\!A={}^{(4)}{A^a}_a=-3e^{-2f}\nabla^2 f.
\ee
\subsection{Calculation of ${}^{(5)}\!A_{ab}$}
The matrix ${}^{(5)}\!A_{ab}={C^m}_{an}{C^n}_{bm}$ is symmetric. Indeed
$${}^{(5)}\!A_{ba}={C^m}_{bn}{C^n}_{am}={C^n}_{bm}{C^m}_{an}={}^{(5)}\!A_{ab}.$$
By straightforward calculations we obtain
\be
\label{B-A5}
{}^{(5)}\!A_{ab}=e^{-2f}\left(
\begin{array}{cccc}
0 &0 &0&0\\
0&  3f_1^2 &3f_1f_2&3 f_1f_3\\   
0& 3f_1f_2 & 3f_2^2 & 3f_2f_3\\
0& 3f_1f_3 & 3f_2f_3 & 3f_3^2
\end{array}
\right)
\ee
The trace of this matrix is
\be
\label{B-A4t}
{}^{(4)}\!A={}^{(5)}{A^a}_a=-3e^{-2f}\nabla^2 f.
\ee
\subsection{Calculation of ${}^{(6)}\!A_{ab}$}
The matrix ${}^{(6)}\!A_{ab}=C_{man}{{C^m}_b}^n$ is symmetric. 
It can be rewritten explicitly as
\brn
 {}^{(6)}\!A_{ab}&=&{C^0}_{a0}{C^0}_{b0}-{C^0}_{a1}{C^0}_{b1}
-{C^0}_{a2}{C^0}_{b2}-{C^0}_{a3}{C^0}_{b3}\\
&&-{C^1}_{a0}{C^1}_{b0}+
{C^1}_{a1}{C^1}_{b1}+{C^1}_{a2}{C^1}_{b2}+{C^1}_{a3}{C^1}_{b3}\\
&&-{C^2}_{a0}{C^2}_{b0}+
{C^2}_{a1}{C^2}_{b1}+{C^2}_{a2}{C^2}_{b2}+{C^2}_{a3}{C^2}_{b3}\\
&&-{C^3}_{a0}{C^3}_{b0}+
{C^3}_{a1}{C^3}_{b1}+{C^3}_{a2}{C^3}_{b2}+{C^3}_{a3}{C^3}_{b3}
\ern
The direct calculations give 
\be
\label{B-A6}
{}^{(6)}\!A_{ab}=e^{-2f}\left(
\begin{array}{cccc}
-\nabla^2f &0 &0&0\\
0&2f_1^2+\nabla^2f &2f_1f_2&2 f_1f_3\\   
0& 2f_1f_2 & 2f_2^2+\nabla^2f & 2f_2f_3\\
0& 2f_1f_3 & 2f_2f_3 & 2f_3^2+\nabla^2f
\end{array}
\right)
\ee
The trace of the matrix is 
\be
\label{B-A5t}
{}^{(6)}{A^a}_a={}^{(2)}\!A=-6e^{-2f}\nabla^2 f.
\ee
\subsection{Calculation of ${}^{(7)}\!A_{ab}$}
The matrix ${}^{(7)}\!A_{ab}=C_aC_b$ is symmetric and its explicit form is
\be
{}^{(7)}\!A_{ab}=e^{-2f}\left(
\begin{array}{cccc}
0 &0 &0&0\\
0&f_1^2 &f_1f_2& f_1f_3\\   
0& f_1f_2 & f_2^2 & f_2f_3\\
0& f_1f_3 & f_2f_3 & f_3^2
\end{array}
\right)
\ee
The trace of this matrix is 
\be
{}^{(6)}\!A={}^{(7)}{A^a}_a={-}^{(1)}\!A=-e^{-2f}\nabla^2 f.
\ee


}
\end{document}